\documentclass
[10pt,twocolumn,showpacs,superscriptaddress,balancelastpage,pra]{revtex4}%
\usepackage{amssymb}
\usepackage{graphicx}
\usepackage{amsmath}
\usepackage{amsfonts}%
\ifx\pdfoutput\relax\let\pdfoutput=\undefined\fi
\newcount\msipdfoutput
\ifx\pdfoutput\undefined\else
\ifcase\pdfoutput\else
\ifx\paperwidth\undefined\else
\ifdim\paperheight=0pt\relax\else\pdfpageheight\paperheight\fi
\ifdim\paperwidth=0pt\relax\else\pdfpagewidth\paperwidth\fi
\fi\fi\fi
\begin{document}
\title{Violations of entropic Bell inequalities with coarse-grained quadrature
measurements for continuous-variable states}
\author{Zeng-Bing Chen}
\email{zbchen@ustc.edu.cn}
\affiliation{Hefei National Laboratory for Physical Sciences at Microscale and Department
of Modern Physics, University of Science and Technology of China, Hefei, Anhui
230026, People's Republic of China}
\affiliation{The CAS Center for Excellence in QIQP and the Synergetic Innovation Center for
QIQP, University of Science and Technology of China, Hefei, Anhui 230026,
People's Republic of China}
\author{Yao Fu}
\email{yaofu@mail.ustc.edu.cn}
\affiliation{Hefei National Laboratory for Physical Sciences at Microscale and Department
of Modern Physics, University of Science and Technology of China, Hefei, Anhui
230026, People's Republic of China}
\affiliation{The CAS Center for Excellence in QIQP and the Synergetic Innovation Center for
QIQP, University of Science and Technology of China, Hefei, Anhui 230026,
People's Republic of China}
\author{Yu-Kang Zhao}
\email{ykzhao@mail.ustc.edu.cn}
\affiliation{Department of Modern Physics, University of Science and Technology of China,
Hefei, Anhui 230026, People's Republic of China}
\date{\today }


\begin{abstract}
It is a long-standing belief, as pointed out by Bell in 1986, that it is
impossible to use a two-mode Gaussian state possessing a positive-definite
Wigner function to demonstrate nonlocality as the Wigner function itself
provides a local hidden-variable model. In particular, when one performs
continuous-variable (CV) quadrature measurements upon a routinely generated CV
entanglement, namely, the two-mode squeezed vacuum (TMSV) state, the resulting
Wigner function is positive-definite and as such, the TMSV state cannot
violate any Bell inequality using CV quadrature measurements. We show here,
however, that a Bell inequality for CV states in terms of entropies can be
quantum mechanically violated by the TMSV state with two
\textit{coarse-grained} quadrature measurements per site within experimentally
accessible parameter regime. The proposed CV entropic Bell inequality is
advantageous for an experimental test, especially for a possible loophole-free
test of nonlocality, as the quadrature measurements can be implemented with
homodyne detections of nearly $100\%$ detection efficiency under current technology.

\end{abstract}
\maketitle

\section{Introduction}

Local realism, as first introduced by Einstein, Podolsky and Rosen (EPR) in
their famous paper \cite{EPR}, is the cornerstone of one's classical
intuitions, namely, physical systems have local \textquotedblleft elements of
reality\textquotedblright\ no matter which experiment actually was performed.
Then, the most radical departure of quantum mechanics from the classical
intuitions is the discovery of Bell's inequalities \cite{Bell,CHSH}, which
enable quantitative tests of quantum mechanics against local realism. While
the original Bell inequalities (or in a broader sense, Bell's theorem) were
derived for discrete quantum variables, their various extensions \cite{RMP}
have been developed for a large number of different settings. Particularly,
the original EPR paradox for continuous variables (CVs) has been a source of
renewed interest for topics such as the preparation of the EPR-type states
\cite{Reid,Ou,Ou-APB,Bvan,EPR-RMP} and nonlocality
\cite{pulsed,Grangier,Banaszek,Chen,phaseAM,LHF-cv,smallV,nonL,moreM}.

For a two-mode Gaussian state possessing a positive-definite Wigner function,
the Wigner function itself provides a local hidden-variable model, and thus it
is impossible to use the state to demonstrate nonlocality. This is a
long-standing belief pointed out by Bell in 1986 \cite{LHF-cv,BellBook}. A
particular example is the two-mode squeezed vacuum (TMSV) state
\cite{Reid,Ou,Ou-APB,Bvan,EPR-RMP}, which is a routinely generated CV
entanglement source and useful for various CV quantum information tasks
\cite{Bvan}. The Wigner function of the TMSV state with CV quadrature
measurements is positive-definite. Therefore, the TMSV state cannot violate
any Bell inequality under the quadrature measurement settings. For this
reason, the existing proposals for the Bell test with CV systems
\cite{pulsed,Grangier,Banaszek,Chen,phaseAM,LHF-cv,smallV,nonL,moreM} make use
of correlation functions for the parity operators \cite{Grangier,Banaszek},
the \textquotedblleft parity-spin\textquotedblright\ operators \cite{Chen}, or
the quantities acquired by a binning process to convert the continuous
outcomes into binary results \cite{phaseAM,LHF-cv,smallV}, enabling us to use the
ordinary form of Bell's inequalities. To take the full capacity of CV nature,
other strategies were proposed with additional mechanisms, like using
non-Gaussian states together with certain (effective) nonlinearity, but with
relatively small violations \cite{LHF-cv,nonL}, or more involved modes
\cite{moreM}.

We show in this paper, however, that a Bell inequality for CV states in terms
of entropies can be quantum mechanically violated by the TMSV state with two
\textit{coarse-grained} quadrature measurements per site using the homodyne
detection technique. Bell's inequalities formulated in an
information-theoretic context \cite{BC,Schumacher,Cerf}, i.e., the entropic
Bell inequalities, were first proposed by Braunstein and Caves and present a
new angle to the conceptually important topic, namely, the quantum violations
of local realism. An appealing feature of the entropic Bell inequalities is
that they are applicable to a pair of $N$-level quantum systems for arbitrary
$N$. Despite this, it is still a nontrivial question on how to demonstrate
the quantum violations of local realism with CV systems, using
information-theoretic Bell inequalities. For the very definition of
information or entropy, there are important differences \cite{infT} between
the CV and discrete-variable cases. To circumvent the difficulty caused by
these differences, the proposed CV entropic Bell inequality uses a pair of
coarse-grained quadrature measurements per site. In this way, only
experimentally measured discrete probability distributions \cite{discA,discL}
are involved. Furthermore, instead of correlation functions as used in the
usual Bell inequalities, we use the conditional or mutual entropies to give
the constraints of local realism. We then demonstrate the quantum mechanical
violations of the CV entropic Bell inequality by the TMSV state without any
use of non-Gaussian states or more involved modes. The CV entropic Bell
inequality as proposed here is friendly to an experimental test, especially
for a possible loophole-free test of nonlocality, as the homodyne detections
have nearly $100\%$ detection efficiency \cite{Hom} under current technology.

\section{The CV entropic Bell inequality}

An important trick in our argument is that, instead of using probability
density $p(a)$ for a continuous random variable $\mathbf{a}$ \cite{infT}, we
use only experimentally measured discrete probability distributions
\cite{discA}, which were proved to be very useful for witnessing CV
entanglement \cite{discL}. As the continuous probability density $p(a)$ cannot
be determined with a finite number of measurements, one can, however, measure
$\mathbf{a}$ to discrete windows $\mathbf{A}_{\ell}$ of size $\Delta a$
(coarse-grained measurements). Namely, the continuous random variable
$\mathbf{a}$ is discretized into equally spaced windows of size $\Delta a$
reflecting the precision of the experimental setup. Then the probability of
measuring $\mathbf{a}$ to be in window $\mathbf{A}_{\ell}$ reads
\cite{discA,discL}
\begin{equation}
P(\mathbf{A}_{\ell})\equiv\int_{\Delta a_{\ell}}dap(a), \label{disP}%
\end{equation}
where the integration is performed over $\Delta a_{\ell}\equiv\lbrack a_{\ell
}-\frac{1}{2}\Delta a,a_{\ell}+\frac{1}{2}\Delta a]$ with $a_{\ell}=\ell
\cdot\Delta a$ ($\ell=0,\pm1,\pm2,...$). The Shannon entropy of this discrete
probability distribution $P(\mathbf{A}_{\ell})$ is given by
\begin{equation}
S(\mathbf{A})=-\sum_{\ell}P(\mathbf{A}_{\ell})\ln P(\mathbf{A}_{\ell}).
\label{SX}%
\end{equation}
The usual \textit{differential} information (or entropy) for the continuous
random variable $\mathbf{a}$ with probability density $p(a)$ reads
\begin{equation}
s(\mathbf{a})=-\int dap(a)\ln p(a). \label{sx}%
\end{equation}
Note that $s(\mathbf{a})$ is defined up to an arbitrary constant and can even
be arbitrarily large, positive or negative, because of the continuous nature
of the random variable $\mathbf{a}$. This property of the CV information is in
contrast to the discrete-variable cases.

For two random variables $\mathbf{a}$ and $\mathbf{b}$ having a joint
probability density $p(a,b)$, we can define correspondingly an information
$s(\mathbf{a,b})=-\int dadbp(a,b)\ln p(a,b)$. The conditional information
reads
\begin{equation}
s(\mathbf{a}\left\vert \mathbf{b})\right.  =-\int dbp(b)\int dap(a\left\vert
b)\right.  \ln p(a\left\vert b)\right.  \label{sxy}%
\end{equation}
in terms of the conditional probability density $p(a\left\vert b)\right.  $.
Using Bayes' theorem, i.e., $p(a,b)=p(a\left\vert b)\right.
p(b)=p(b\left\vert a)\right.  p(a)$, one has
\begin{equation}
s(\mathbf{a},\mathbf{b})=s(\mathbf{a}\left\vert \mathbf{b})\right.
+s(\mathbf{b})=s(\mathbf{b}\left\vert \mathbf{a})\right.  +s(\mathbf{a}).
\label{ssss}%
\end{equation}
Similarly, one can define the Shannon entropy $S(\mathbf{A},\mathbf{B})$ of
the discrete probability distribution $P(\mathbf{A}_{\ell},\mathbf{B}_{m})$.
The discretized conditional entropy is then
\begin{equation}
S(\mathbf{A}\left\vert \mathbf{B})\right.  =S(\mathbf{A},\mathbf{B}%
)-S(\mathbf{B}). \label{SXY}%
\end{equation}

In terms of the above discretized entropies we have two useful inequalities
\begin{equation}
S(\mathbf{A},\mathbf{B})\geq S(\mathbf{A})\geq S(\mathbf{A}\left\vert
\mathbf{B})\right.  , \label{sss}%
\end{equation}
which have a transparent information-theoretic interpretation. The first
inequality stems from the fact that the information carried by two quantities
is never less than the information separately carried by either quantity. The
information carried by a quantity never decreases by removing a condition.
This then leads to the last inequality. Note that for the CV case, one does
not have the inequalities $s(\mathbf{a},\mathbf{b})\geq s(\mathbf{a})\geq
s(\mathbf{a}\left\vert \mathbf{b})\right.  $.

To derive the required entropic Bell inequality we follow the reasoning by
Braunstein and Caves \cite{BC}. Consider now two space-like separated CV
quantum systems $A$ and $B$. For system $A$ ($B$) we have two measurable
quantities $\mathbf{a}$ and $\mathbf{a}^{\prime}$ ($\mathbf{b}$ and
$\mathbf{b}^{\prime}$) whose values are denoted by continuous random variables
$a$ and $a^{\prime}$ ($b$ and $b^{\prime}$). Quantum mechanically, two
incompatible observables (e.g., $\mathbf{a}$ and $\mathbf{a}^{\prime}$ for
system $A$) of a system cannot be measured simultaneously. Hence in each run
of the two-setting Bell experiments one can only measure two observables
(here, e.g., $\mathbf{a}$ and $\mathbf{b}$), one from each system. By
contrast, local realism implies that the four quantities specified above are
all local objective properties of the whole system. An important consequence
of the observation is the existence of a joint probability density
$p(a,a^{\prime},b,b^{\prime})$, from which we can obtain appropriate
probability densities, e.g.,
\begin{equation}
p(a,b)=\int da^{\prime}db^{\prime}p(a,a^{\prime},b,b^{\prime}), \label{pab}%
\end{equation}
the marginals of the joint probability density $p(a,a^{\prime},b,b^{\prime})$.
Then other relevant probability densities can also be obtained, e.g.,
$p(a)=\int dap(a,b)$.

With these probability densities in mind, we can define the corresponding
discrete entropies, in terms of which we have the following information
inequality \cite{BC}
\begin{equation}
S(\mathbf{A},\mathbf{B})\leq S(\mathbf{A},\mathbf{B}^{\prime},\mathbf{A}%
^{\prime},\mathbf{B}), \label{sabab}%
\end{equation}
where $S(\mathbf{A},\mathbf{B}^{\prime},\mathbf{A}^{\prime},\mathbf{B}%
)=S(\mathbf{A}\left\vert \mathbf{B}^{\prime},\mathbf{A}^{\prime}%
,\mathbf{B})\right.  +S(\mathbf{B}^{\prime}\left\vert \mathbf{A}^{\prime
},\mathbf{B})\right.  +S(\mathbf{A}^{\prime}\left\vert \mathbf{B})\right.
+S(\mathbf{B})$. Using the facts that $S(\mathbf{A}\left\vert \mathbf{B}%
^{\prime},\mathbf{A}^{\prime},\mathbf{B})\right.  \leq S(\mathbf{A}\left\vert
\mathbf{B}^{\prime})\right.  $, $S(\mathbf{B}^{\prime}\left\vert
\mathbf{A}^{\prime},\mathbf{B})\right.  \leq S(\mathbf{B}^{\prime}\left\vert
\mathbf{A}^{\prime})\right.  $ and Eq.~(\ref{SXY}), we have the entropic Bell
inequality
\begin{equation}
0\leq S(\mathbf{A}\left\vert \mathbf{B}^{\prime})\right.  +S(\mathbf{B}%
^{\prime}\left\vert \mathbf{A}^{\prime})\right.  +S(\mathbf{A}^{\prime
}\left\vert \mathbf{B})\right.  -S(\mathbf{A}\left\vert \mathbf{B})\right.  .
\label{eBell}%
\end{equation}
This is an information-theoretic constraint that has to be obeyed by all local
realistic theories. Similar to Ref.~\cite{Cerf} we can also obtain an
entropic Bell inequality in terms of the mean mutual information
$S(\mathbf{A};\mathbf{B}^{\prime})+S(\mathbf{A}^{\prime};\mathbf{B}^{\prime
})+S(\mathbf{A}^{\prime};\mathbf{B})-S(\mathbf{A};\mathbf{B})\leq
S(\mathbf{A}^{\prime})+S(\mathbf{B}^{\prime})$, which takes a form quite
similar to the Clauser-Horne-Shimony-Holt inequalities \cite{CHSH}. Here
$S(\mathbf{A};\mathbf{B})=S(\mathbf{A})+S(\mathbf{B})-S(\mathbf{A}%
,\mathbf{B})$.

\section{Quantum violation of the CV entropic Bell inequality}

Now let us show the quantum violation of the entropic Bell inequality
Eq. (\ref{eBell}) by the regularized EPR states \cite{Reid,Ou,Ou-APB,pulsed}
produced in a pulsed nondegenerate optical parametric amplification process.
The process generates the TMSV state\ associated with two quantized light
modes (denoted by the corresponding annihilation operators $\hat{a}$ and
$\hat{b}$) as
\begin{align}
\left\vert \mathrm{TMSV}\right\rangle  &  =e^{r(\hat{a}^{\dagger}\hat
{b}^{\dagger}-\hat{a}\hat{b})}|00\rangle=\frac{1}{\cosh r}e^{\hat{a}^{\dagger
}\hat{b}^{\dagger}\tanh r}|00\rangle\nonumber\\
&  =\frac{1}{\cosh r}\sum_{n=0}^{\infty}(\tanh r)^{n}|n,n\rangle. \label{NOPA}%
\end{align}
Here, for simplicity, we assume that $r>0$ known as the squeezing parameter and
$\left\vert nn\right\rangle \equiv\left\vert n\right\rangle _{a}\left\vert
n\right\rangle _{b}=\frac{1}{n!}\hat{a}^{\dagger}{}^{n}\hat{b}^{\dagger}{}%
^{n}\left\vert 00\right\rangle $. In the infinite squeezing limit, the TMSV
state $\left\vert \mathrm{TMSV}\right\rangle $ becomes the original,
normalized EPR state \cite{Ou-APB,Chen}; i.e., $\left\vert \mathrm{TMSV}%
\right\rangle \overset{r\rightarrow\infty}{\longrightarrow}\left\vert
\mathrm{EPR}\right\rangle _{\mathrm{normalized}}$. Such a CV entanglement can
now be routinely generated and is a vital resource for various CV quantum
information tasks \cite{Bvan}.

For each mode of the light field, one measures the quadrature phase amplitude
operators
\begin{equation}
\mathbf{\hat{a}}_{\theta}=\tfrac{1}{\sqrt{2}}(\hat{a}^{\dagger}e^{i\theta
}+\hat{a}e^{-i\theta}),\text{ \ }\mathbf{\hat{b}}_{\phi}=\tfrac{1}{\sqrt{2}%
}(\hat{b}^{\dagger}e^{i\phi}+\hat{b}e^{-i\phi}), \label{ab}%
\end{equation}
with the usual homodyne measurement technique. Note that $\left[
\mathbf{\hat{a}}_{\theta},\mathbf{\hat{a}}_{\theta+\pi/2}\right]  =i$. Thus,
$\mathbf{\hat{a}}_{\theta}$ and $\mathbf{\hat{a}}_{\theta+\pi/2}$ (similarly
for $\mathbf{\hat{b}}_{\phi}$ and $\mathbf{\hat{b}}_{\phi+\pi/2}$) form a
canonically conjugate pair. Experimental control of the local oscillator
phases ($\theta$ and $\phi$) provides access to the continuous distribution of
the quadratures $\mathbf{\hat{a}}_{\theta}$ and $\mathbf{\hat{b}}_{\phi}$. Let
us denote the eigenvectors of $\mathbf{\hat{a}}_{\theta}$ and $\mathbf{\hat
{b}}_{\phi}$ by $\left\vert a_{\theta}\right\rangle $ and $\left\vert b_{\phi
}\right\rangle $, respectively, namely, $\mathbf{\hat{a}}_{\theta}\left\vert
a_{\theta}\right\rangle =a_{\theta}\left\vert a_{\theta}\right\rangle $ and
$\mathbf{\hat{b}}_{\phi}\left\vert b_{\phi}\right\rangle =b_{\phi}\left\vert
b_{\phi}\right\rangle $. Then the quantum pair probability density for
measuring $\mathbf{\hat{a}}_{\theta}$ and $\mathbf{\hat{b}}_{\phi}$ (with the
results $a_{\theta}$ and $b_{\phi}$) upon $\left\vert \mathrm{TMSV}%
\right\rangle $ can be calculated as
\begin{equation}
p_{\mathrm{QM}}(a_{\theta},b_{\phi})=\left\vert \left\langle a_{\theta
}\right\vert \left\langle b_{\phi}\right.  \left\vert \mathrm{TMSV}%
\right\rangle \right\vert ^{2}. \label{abNOPA}%
\end{equation}

To evaluate $p_{\mathrm{QM}}(a_{\theta},b_{\phi})$, we need some useful
properties \cite{QOptics} of $\left\vert a_{\theta}\right\rangle $\ (similarly
for $\left\vert b_{\phi}\right\rangle $), namely,
\begin{equation}
\left.  \left\langle a_{\theta}\right.  \left\vert a_{\theta}^{\prime
}\right\rangle =\delta(a_{\theta}-a_{\theta}^{\prime})\right.  ,\text{
\ \ }\left.  \int da_{\theta}\left\vert a_{\theta}\right\rangle \left\langle
a_{\theta}\right\vert =\hat{I}_{A}\right.  , \label{oc}%
\end{equation}
which are the orthogonal and completeness relations for $\left\vert a_{\theta
}\right\rangle $, similar to the usual eigenvectors of a position operator.
Here, $\hat{I}_{A}$ is the identity operator for system $A$. Moreover,
\begin{align}
\left\langle a_{\theta}\right.  \left\vert n\right\rangle _{a}  &  =\frac
{1}{\sqrt{\sqrt{\pi}2^{n}n!}}\exp(-in\theta-a_{\theta}^{2}/2)H_{n}(a_{\theta
}),\nonumber\\
\left\langle b_{\phi}\right.  \left\vert n\right\rangle _{b}  &  =\frac
{1}{\sqrt{\sqrt{\pi}2^{n}n!}}\exp(-in\phi-b_{\phi}^{2}/2)H_{n}(b_{\phi}),
\label{Herm-a}%
\end{align}
where $H_{n}$ is the Hermite polynomial of order $n$. Using Eq.~(\ref{Herm-a})
we can write down $\left\vert a_{\theta}\right\rangle $ explicitly
\cite{QOptics}
\begin{equation}
\left\vert a_{\theta}\right\rangle =\pi^{-1/4}\exp\left[  -\tfrac{1}%
{2}a_{\theta}^{2}+\sqrt{2}e^{i\theta}a_{\theta}\hat{a}^{\dag}-\tfrac{1}%
{2}e^{2i\theta}\hat{a}^{\dag2}\right]  \left\vert 0\right\rangle .
\label{explicit}%
\end{equation}
Hence, one can evaluate $p_{\mathrm{QM}}(a_{\theta},b_{\phi})$ using either
Eq.~(\ref{Herm-a}) or Eq.~(\ref{explicit}).

Here we calculate $p_{\mathrm{QM}}(a_{\theta},b_{\phi})$ with the help of
Eq.~(\ref{Herm-a}). To begin with, we use Mehler's formula \cite{specialF},
\begin{equation}
\sum_{n=0}^{+\infty}\frac{H_{n}(x)H_{n}(y)}{2^{n}n!}t^{n}=\frac{1}%
{\sqrt{1-t^{2}}}e^{[2xyt-(x^{2}+y^{2})t^{2}]/(1-t^{2})}, \label{Meh}%
\end{equation}
where $\left\vert t\right\vert <1$. Consequently, using Eqs.~(\ref{NOPA}) and
(\ref{Herm-a}) yields
\begin{equation}
\left\langle a_{\theta}\right\vert \left\langle b_{\phi}\right.  \left\vert
\mathrm{TMSV}\right\rangle =\frac{e^{-(a_{\theta}^{2}+b_{\phi}^{2}%
)/2}e^{[2a_{\theta}b_{\phi}t-(a_{\theta}^{2}+b_{\phi}^{2})t^{2}]/(1-t^{2})}%
}{\sqrt{\pi}\sqrt{1-t^{2}}\cosh r}, \label{inner}%
\end{equation}
with $t=e^{-i(\theta+\phi)}\tanh r$ (Note that indeed $\left\vert t\right\vert
<1$ as $\tanh r<1$ for $r>0$). Hereafter, we take $\varphi=\theta+\phi$. Then
\begin{align}
p_{\mathrm{QM}}(a_{\theta},b_{\phi})  &  =\frac{e^{-(a_{\theta}^{2}+b_{\phi
}^{2})}\left\vert e^{[2a_{\theta}b_{\phi}t-(a_{\theta}^{2}+b_{\phi}^{2}%
)t^{2}]/(1-t^{2})}\right\vert ^{2}}{\pi\left\vert 1-t^{2}\right\vert \cosh
^{2}r}\nonumber\\
&  =\frac{e^{-(a_{\theta}^{2}+b_{\phi}^{2})v+2a_{\theta}b_{\phi}w}}%
{\pi\left\vert 1-t^{2}\right\vert \cosh^{2}r}, \label{pqm}%
\end{align}
which shows the explicit anti-correlations between the local oscillator phases
$\theta$ and $\phi$. Here $\left\vert 1-t^{2}\right\vert =\sqrt{1+\tanh
^{4}r-2\tanh^{2}r\cos2\varphi}$ and
\begin{align}
v  &  =\frac{1-\tanh^{4}r}{\left\vert 1-t^{2}\right\vert ^{2}},\nonumber\\
w  &  =\frac{2\tanh r\cos\varphi(1-\tanh^{2}r)}{\left\vert 1-t^{2}\right\vert
^{2}}. \label{vw}%
\end{align}

From the Gaussian integral formula
\begin{equation}
\int\exp(-%
{\textstyle\sum\nolimits_{i,j=1}^{m}}
x_{i}\mathcal{A}_{ij}x_{j})d^{m}x=\sqrt{\frac{\pi^{m}}{\det\mathcal{A}}},
\label{gaussI}%
\end{equation}
we obtain in particular that $\int dxdy\exp[-(x^{2}+y^{2})\alpha+2xy\beta
]=\pi/\sqrt{\alpha^{2}-\beta^{2}}\equiv J(\alpha,\beta)$, where $\alpha>0$.
Here, $\mathcal{A}_{m\times m}$ is a symmetric positive-definite (hence
invertible) covariance matrix. Then we can confirm the normalization condition
of $p_{\mathrm{QM}}(a_{\theta},b_{\phi})$,
\begin{equation}
\int da_{\theta}db_{\phi}p_{\mathrm{QM}}(a_{\theta},b_{\phi})=\frac{\pi
/\sqrt{v^{2}-w^{2}}}{\pi\left\vert 1-t^{2}\right\vert \cosh^{2}r}=1,
\label{gyh}%
\end{equation}
as can easily be checked by using Eq.~(\ref{vw}). From $p_{\mathrm{QM}%
}(a_{\theta},b_{\phi})$ we can easily obtain
\begin{equation}
p_{\mathrm{QM}}(b_{\phi})=\int da_{\theta}p_{\mathrm{QM}}(a_{\theta},b_{\phi
})=\frac{e^{-b_{\phi}^{2}/\cosh2r}}{\sqrt{\pi\cosh2r}}. \label{paqm}%
\end{equation}

Now let us calculate $S_{\mathrm{QM}}(\mathbf{B}_{\phi})$ and $S_{\mathrm{QM}%
}(\mathbf{A}_{\theta},\mathbf{B}_{\phi})$. Hereafter, we take base of the
logarithm to be $e$. By definition, $S_{\mathrm{QM}}(\mathbf{B}_{\phi}%
)=-\sum_{m}P_{\mathrm{QM}}(\mathbf{B}_{\phi,m})\ln P_{\mathrm{QM}}%
(\mathbf{B}_{\phi,m})$ and $S_{\mathrm{QM}}(\mathbf{A}_{\theta},\mathbf{B}%
_{\phi})=-\sum_{\ell,m}P_{\mathrm{QM}}(\mathbf{A}_{\theta,\ell},\mathbf{B}%
_{\phi,m})\ln P_{\mathrm{QM}}(\mathbf{A}_{\theta,\ell},\mathbf{B}_{\phi,m})$,
where
\begin{align}
P_{\mathrm{QM}}(\mathbf{B}_{\phi,m})  &  =\int_{\Delta b_{\phi,m}}db_{\phi
}p_{\mathrm{QM}}(b_{\phi}),\nonumber\\
P_{\mathrm{QM}}(\mathbf{A}_{\theta,\ell},\mathbf{B}_{\phi,m})  &
=\int_{\Delta a_{\theta,\ell}}da_{\theta}\int_{\Delta b_{\phi,m}}db_{\phi
}p_{\mathrm{QM}}(a_{\theta},b_{\phi}). \label{sbsab}%
\end{align}
With these results in mind, we get the quantum discretized conditional entropy
under the given measurement precisions ($\Delta a_{\theta}$ and $\Delta
b_{\phi}$):
\begin{equation}
S_{\mathrm{QM}}(\mathbf{A}_{\theta}\left\vert \mathbf{B}_{\phi})\right.
=S_{\mathrm{QM}}(\mathbf{A}_{\theta},\mathbf{B}_{\phi})-S_{\mathrm{QM}%
}(\mathbf{B}_{\phi})\equiv S_{\mathrm{QM}}(\varphi). \label{ssqm}%
\end{equation}
The negativity of $S_{\mathrm{QM}}(\mathbf{A}_{\theta}\left\vert
\mathbf{B}_{\phi})\right.  $ rules out any description of local realism, or an
underlying joint probability distribution \cite{Cerf}.%

\begin{figure}[tbh]
\centering \resizebox{9cm}{!}{\includegraphics{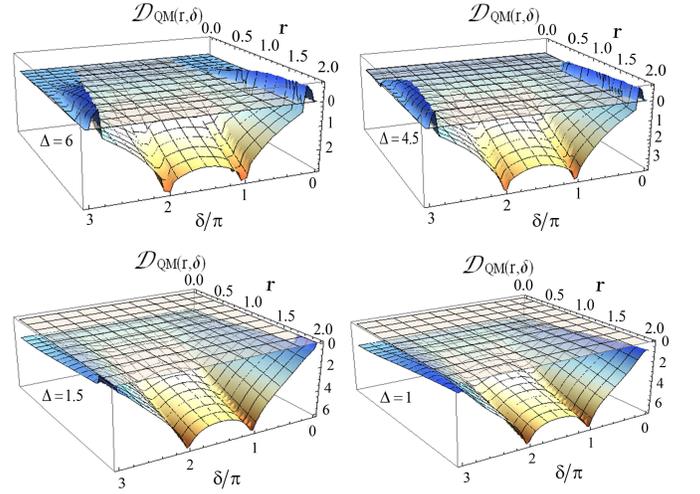}}
\caption{(Color online). The plot of $\mathcal{D}_{\mathrm{QM}}$ [see
Eq.~(\ref{sqm})] as a function of angle $\delta$ and the squeezing parameter
$r$ for different values of the measurement resolutions $\Delta$. We also show
the plane of $\mathcal{D}_{\mathrm{QM}}=0$, above which is the parameter
regime of violating the inequality Eq. (\ref{eBell}). An apparent trend is that
larger $\Delta$ requires smaller $r$ to violate the entropic Bell inequality.}
\label{fig}
\end{figure}

To demonstrate that the entropic Bell inequality Eq. (\ref{eBell}) can indeed be
violated quantum mechanically, we take the following angles:
\begin{equation}
\phi=-\theta+\delta,\text{ \ }\theta^{\prime}=\theta-2\delta/3,\text{ \ }%
\phi^{\prime}=-\theta+\delta/3. \label{angle}%
\end{equation}
Then the right-hand side of Eq.~(\ref{eBell}) becomes [Note that
$S_{\mathrm{QM}}(-\varphi)=S_{\mathrm{QM}}(\varphi)$]
\begin{equation}
\mathcal{D}_{\mathrm{QM}}(r,\delta)=3S_{\mathrm{QM}}(\frac{\delta}%
{3})-S_{\mathrm{QM}}(\delta). \label{sqm}%
\end{equation}
Here, we assumed that $\Delta a_{\theta}=\Delta a_{\theta^{\prime}}\equiv\Delta
a$ and $\Delta b_{\phi}=\Delta b_{\phi^{\prime}}\equiv\Delta b$. The
negativity of $\mathcal{D}_{\mathrm{QM}}$ implies the quantum violations of
the entropic Bell inequality Eq. (\ref{eBell}) and gives the deficit information
\cite{BC} carried by systems $A$ and $B$, relative to that imposed by local
realism under the same geometry.

\begin{figure}
[ptb]
\begin{center}
\ifcase\msipdfoutput
\includegraphics[
height=1.7793in,
width=3.7218in
]%
{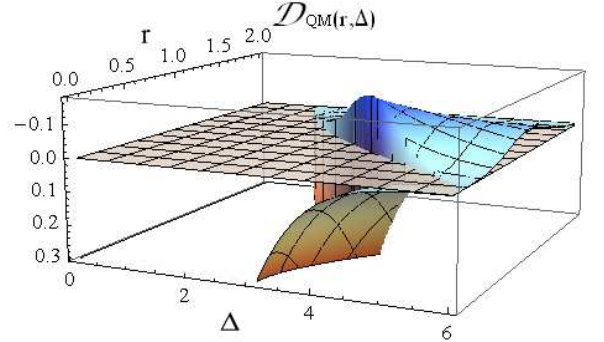}%
\else
\includegraphics[
height=1.7793in,
width=3.7218in
]%
{fig2.eps}%
\fi
\caption{(Color online). The plot of $\mathcal{D}_{\mathrm{QM}}$ at $\delta=0$
as a function of the squeezing parameter $r$ and the measurement resolution
$\Delta$. The parameter regime violating the inequality Eq. (\ref{eBell}) is above
the plane of $\mathcal{D}_{\mathrm{QM}}(r,\Delta,\delta=0)=0$.}%
\label{figdrd}%
\end{center}
\end{figure}

In Fig.~\ref{fig} we plot $\mathcal{D}_{\mathrm{QM}}$ as a function of
$\delta$ and $r$ for different values of $\Delta a=\Delta b\equiv\Delta$. We
also show the plane of $\mathcal{D}_{\mathrm{QM}}=0$, above which is the
parameter regime of violating the inequality Eq. (\ref{eBell}). Numeric result at
$\Delta=1.5$ shows only tiny violation for large squeezing parameters in the
vicinity of $r=2$, while no violation was found for $\Delta=1$ and
$r\in\lbrack0,2]$. From Fig.~\ref{fig}, we note that a large fraction of the
violations occur in the vicinity of $\delta=0$ (as well as its symmetric point
$\delta=3\pi$). We then plot $\mathcal{D}_{\mathrm{QM}}$ at $\delta=0$ as a
function of $\Delta$ and $r$ in Fig.~\ref{figdrd}, where the plane of
$\mathcal{D}_{\mathrm{QM}}(r,\Delta,\delta=0)=0$ is shown, too. In this
specific case the parameter regime of violating the entropic Bell inequality
indicates that larger values of $\Delta$ require smaller squeezing parameters
$r$, a similar trend observed from Fig.~\ref{fig}. Thus, the proposed
violations are experimentally more accessible for coarse-grained measurements
with larger $\Delta$.%

However, if we only consider the maximal violations of the inequality Eq.
(\ref{eBell}), namely, the minimal value $\mathcal{D}_{\mathrm{QM}}^{\min}$ of
$\mathcal{D}_{\mathrm{QM}}(r,\Delta,\delta)$ for a given parameter regime, a
complicated interplay between $r$ and $\Delta$ is observed during our numeric
calculations. For example, one can detect two minimal values of $\mathcal{D}%
_{\mathrm{QM}}(r,\Delta,\delta)$ for the parameter regime of $r\in\lbrack
0,2]$, $\delta\in\lbrack0,\pi]$, and small $\Delta$ ($\leq10$); i.e.,
$\mathcal{D}_{\mathrm{QM}}^{\min}\approx-0.75$ at $r\approx1.817$, $\Delta=6$,
$\delta\approx0.213\pi$ and $r\approx1.915$, $\Delta=3.5$, $\delta
\approx0.098\pi$, respectively. We also calculate $\mathcal{D}_{\mathrm{QM}%
}^{\min}$ for larger $\Delta$. For $\Delta=30$, $\mathcal{D}_{\mathrm{QM}%
}^{\min}\approx-0.0016$ ($r\approx1.991$, $\delta\approx0.502\pi$) for
$r\in\lbrack0,2]$, $-0.685$ ($r\approx3$, $\delta\approx0.335\pi$) for
$r\in\lbrack0,3]$, $-1$ ($r\approx3.756$, $\delta\approx0.0004\pi$) for
$r\in\lbrack0,4]$, indicating more violations for larger $r$ with the given
$\Delta$. For the parameter regime of $r\in\lbrack0,4]$ and $\delta\in
\lbrack0,\pi]$ with even larger $\Delta$, we show two examples, namely,
$\Delta=50$, $\mathcal{D}_{\mathrm{QM}}^{\min}\approx-1.9$ ($r\approx3.781$,
$\delta\approx0.0004\pi$), while $\mathcal{D}_{\mathrm{QM}}^{\min}%
\approx-0.549$ ($r\approx3.901$, $\delta\approx0.0065\pi$) for $\Delta=100$.
From all these numeric calculations, we see that the maximal violation of the
inequality Eq. (\ref{eBell}) is strongly influenced by the parameters of $r$ and
$\Delta$, as well as their interplay. Intuitively, one might expect that our
coarse-grained measurement effectively becomes the usual discrete measurements
for sufficiently large $\Delta$. However, it is quite difficult to get a
simple picture due to the complicated interplay between $r$ and $\Delta$ as we
observed in our numeric calculations.

Finally, let us briefly discuss the remaining experimental issues regarding
the test of the entropic Bell inequality Eq. (\ref{eBell}). For given $r$ and
$\delta$, the joint probability distribution $P_{\mathrm{QM}}(\mathbf{A}%
_{\theta,\ell},\mathbf{B}_{\phi,m})$\ can be experimentally measured
\cite{discL,SMM} from the homodyne detection data, giving rise to
$P_{\mathrm{QM}}(\mathbf{B}_{\phi,m})$ as marginal probability distribution.
For experimentally closing the locality loophole, it is important to use
random and fast switching of the local measurement settings such that the
space-time separated measurements are guaranteed. All previous Bell
experiments could in principle be challenged by the detection-efficiency
and/or locality loopholes \cite{Aspect}. The former loophole has been closed
in Ref.~\cite{detection} by using entangled ions, which can be detected with
nearly perfect efficiency. The entangled photons are ideal for closing the
locality loophole, as pioneered by Aspect \textit{et al}. \cite{Aspect82} and
then by Weihs \textit{et al}. \cite{Weihs}. Our proposal has used the pulsed
\cite{cosh10} CV entangled source (being easy to close the locality loophole)
and the quadrature measurement with homodyne detections (having nearly unit
detection efficiency), thus opening up the exciting possibility for a
loophole-free test of local realism against quantum mechanics.

\section{Conclusions and discussions}

In summary, an entropic Bell inequality has been proposed for CV states. Our
argument requires simply coarse-grained quadrature measurements per site,
without any need of experimentally complicated mechanisms such as non-Gaussian
states and nonlinearity \cite{LHF-cv,nonL} or more involved modes
\cite{moreM}. Thus, this is the simplest nonlocality argument for CV systems
with minimal experimental settings. We then demonstrate the quantum violations
of the CV entropic Bell inequality for TMSV states, although a previous belief
claims no violation of any Bell inequality in its ordinary form for the same
TMSV states using CV quadrature measurements. The parameters required for the
violations are well within the experimentally accessible regime. By taking the
full merits of the CV entangled light fields, our argument thus opens up a
strong possibility for a loophole-free test of local realism.

\begin{acknowledgments}
We thank Wen-Fei Cao for his help in numerical calculations. This work was
supported by the National Natural Science Foundation of China under Grant No. 61125502, the Chinese Academy of Sciences, the National
High Technology Research and Development Program of China, and the National
Fundamental Research Program under Grant No. 2011CB921300.
\end{acknowledgments}


\begin{thebibliography}{99}                                                                                               %



\bibitem {EPR}A. Einstein, B. Podolsky, and N. Rosen, Phys. Rev. \textbf{47},
777 (1935).

\bibitem {Bell}J.S. Bell, Physics (Long Island) \textbf{1,} 195 (1964).

\bibitem {CHSH}J.F. Clauser, M.A. Horne, A. Shimony, and R.A. Holt, Phys. Rev.
Lett. \textbf{23,} 880 (1969).

\bibitem {RMP}J.-W. Pan, Z.-B. Chen, C.-Y. Lu, H. Weinfurter, A. Zeilinger,
and M. \.{Z}ukowski, Rev. Mod. Phys. \textbf{84}, 777 (2012), and references therein.

\bibitem {Reid}M.D. Reid and P.D. Drummond, Phys. Rev. Lett. \textbf{60,} 2731
(1988); M.D. Reid, Phys. Rev. A\textbf{\ 40,} 913 (1989).

\bibitem {Ou}Z.Y. Ou, S.F. Pereira, H.J. Kimble, and K.C. Peng, Phys. Rev.
Lett. \textbf{68} 3663 (1992).

\bibitem {Ou-APB}Z.Y. Ou, S.F. Pereira, and H.J. Kimble, Appl. Phys. B: Photophys. Laser Chem. 
\textbf{55}, 265 (1992).

\bibitem {Bvan}S.L. Braunstein and P. van Loock, Rev. Mod. Phys. \textbf{77},
513 (2005).

\bibitem {EPR-RMP}M.D. Reid, P.D. Drummond, W.P. Bowen, E.G. Cavalcanti, P.K.
Lam, H.A. Bachor, U.L. Andersen, and G. Leuchs, Rev. Mod. Phys. \textbf{81},
1727 (2009).

\bibitem {Grangier}P. Grangier, M.J. Potasek, and B. Yurke, Phys. Rev.
A\textbf{\ 38,} 3132 (1988).

\bibitem {Banaszek}K. Banaszek and K. W\'{o}dkiewicz, Phys. Rev.
A\textbf{\ 58}, 4345 (1998); Phys. Rev. Lett. \textbf{82,} 2009 (1999).

\bibitem {pulsed}A. Kuzmich, I.A. Walmsley, and L. Mandel, Phys. Rev. Lett.
\textbf{85}, 1349 (2000).

\bibitem {Chen}Z.-B. Chen, J.-W. Pan, G. Hou, and Y.-D. Zhang, Phys. Rev.
Lett. \textbf{88}, 040406 (2002); A.F. Abouraddy, T. Yarnall, B.E.A. Saleh, and
M.C. Teich, Phys. Rev. A \textbf{75}, 052114 (2007); T. Yarnall, A.F.
Abouraddy, B.E.A. Saleh, and M.C. Teich, Phys. Rev. Lett. \textbf{99}, 170408 (2007).

\bibitem {phaseAM}A. Gilchrist, P. Deuar, and M.D. Reid, Phys. Rev. Lett.
\textbf{80}, 3169 (1998).

\bibitem {LHF-cv}H. Nha and H.J. Carmichael, Phys. Rev. Lett. \textbf{93},
020401 (2004).

\bibitem {smallV}R. Garc\'{\i}a-Patr\'{o}n, J. Fiur\'{a}\v{s}ek, N.J. Cerf, J.
Wenger, R. Tualle-Brouri, and Ph. Grangier, Phys. Rev. Lett. \textbf{93},
130409 (2004); R. Garc\'{\i}a-Patr\'{o}n, J. Fiur\'{a}\v{s}ek, and N.J. Cerf,
Phys. Rev. A \textbf{71}, 022105 (2005).

\bibitem {nonL}M. Paternostro, H. Jeong, and T.C. Ralph, Phys. Rev. A
\textbf{79}, 012101 (2009).

\bibitem {moreM}E.G. Cavalcanti, C.J. Foster, M.D. Reid, and P.D. Drummond,
Phys. Rev. Lett. \textbf{99}, 210405 (2007); A. Salles, D. Cavalcanti, and A.
Ac\'{\i}n, Phys. Rev. Lett. \textbf{101}, 040404 (2008); Q.Y. He, E.G.
Cavalcanti, M.D. Reid, and P.D. Drummond, Phys. Rev. Lett. \textbf{103},
180402 (2009).

\bibitem {BellBook}J.S. Bell, \textit{Speakable and Unspeakable in Quantum
Mechanics} (Cambridge University Press, Cambridge, 1987), Chap. 21.



\bibitem {BC}S.L. Braunstein and C.M. Caves, Phys. Rev. Lett. \textbf{61,} 662
(1988); Ann. Phys. (NY) \textbf{202}, 22 (1990).

\bibitem {Schumacher}B.W. Schumacher, Phys. Rev. A \textbf{44}, 7047 (1991).

\bibitem {Cerf}N.J. Cerf and C. Adami, Phys. Rev. A \textbf{55}, 3371 (1997).

\bibitem {infT}T.M. Cover and J.A. Thomas, \textit{Elements of Information
Theory} (John Wiley \& Sons, Inc., Hoboken, New Jersey, 2006).

\bibitem {discA}\L . Rudnicki, S.P. Walborn, and F. Toscano, Phys. Rev. A
\textbf{85}, 042115 (2012).

\bibitem {discL}J. Schneeloch, P. B. Dixon, G.A. Howland, C.J. Broadbent, and
J.C. Howell, Phys. Rev. Lett. \textbf{110,} 130407 (2013); D.S. Tasca, \L .
Rudnicki, R.M. Gomes, F. Toscano, and S.P. Walborn, Phys. Rev. Lett.
\textbf{110,} 210502 (2013).

\bibitem {Hom}T.C. Zhang, K.W. Goh, C.W. Chou, P. Lodahl, and H.J. Kimble,
Phys. Rev. A \textbf{67}, 033802 (2003); S. Suzuki, H. Yonezawa, F. Kannari,
M. Sasaki, and A. Furusawa, Appl. Phys. Lett. \textbf{89}, 061116 (2006).

\bibitem {QOptics}S.M. Barnett and P.M. Radmore, \textit{Methods in
Theoretical Quantum Optics} (Oxford Uniersity Press, Oxford, 1997).

\bibitem {specialF}G.E. Andrews, R. Askey, and R. Roy, \textit{Special
Functions} (Cambridge University Press, Cambridge, 2000).

\bibitem {SMM}For the experimental measurement of quadrature operator's
probability distribution for a single-mode light field, see D.T. Smithey, M.
Beck, M.G. Raymer, and A. Faridani, Phys. Rev. Lett. \textbf{70}, 1244 (1993).

\bibitem {Aspect}A. Aspect, Nature (London) \textbf{398}, 189 (1999).

\bibitem {detection}M.A. Rowe, D. Kielpinski, V. Meyer, C.A. Sackett, W. M.
Itano, C. Monroe, and D.J. Wineland, Nature (London) \textbf{409}, 791 (2001).

\bibitem {Aspect82}A. Aspect, J. Dalibard, and G. Roger, Phys. Rev. Lett.
\textbf{49}, 1804 (1982).

\bibitem {Weihs}G. Weihs, T. Jennewein, C. Simon, H. Weinfurter, and A.
Zeilinger, Phys. Rev. Lett. \textbf{81}, 5039 (1998).

\bibitem {cosh10}O. Ayt\"{u}r and P. Kumar, Phys. Rev. Lett. \textbf{65}, 1551 (1990).
\end{thebibliography}
\end{document}